\documentstyle[aps,prl,preprint]{revtex}
\begin{document}
\draft
\preprint{BARI-TH 371/99}
\date{January 2000}
\title{
PHASE ORDERING IN CHAOTIC MAP LATTICES WITH ADDITIVE NOISE}
\author{
Leonardo Angelini, Mario Pellicoro, and Sebastiano Stramaglia}
\address{
Dipartimento Interateneo di Fisica\\
Istituto Nazionale di Fisica Nucleare, Sezione di Bari\\
via Amendola 173, 70126 Bari, Italy}
\maketitle
\begin{abstract}
We present some result about phase separation in coupled map lattices with 
additive noise. We show that additive noise acts as an ordering agent in this 
class of systems. In particular, in the weak coupling region, a suitable 
quantity of noise leads to complete ordering. Extrapolating our results at small 
coupling, we deduce that this phenomenon could take place also in the limit of 
zero coupling.
\end{abstract}

\section{Introduction}

Non trivial collective behavior (NTCB) is an interesting feature peculiar to 
extensively chaotic dynamical systems, where the temporal evolution of spatially 
averaged quantities reveals the onset of long-range order in spite of the local 
disorder \cite{chate}\cite{gallas}\cite{chate2}. A simple way to observe NTCB is 
to study models of spatially extended chaotic systems such as coupled map 
lattices (CMLs) that consist of chaotic maps locally coupled diffusively with 
some coupling strength $g$ \cite{chate2}. In these systems one observes 
multistability that is the remainder, for small couplings, of the completely 
uncoupled case \cite{mckay}. For large enough couplings NTCB is observed, 
corresponding to a {\it macroscopic} attractor, well-defined in the 
infinite-size limit and reached for almost every initial condition.

In a recent paper \cite{lemaitre} phase separation mechanisms have been 
investigated in a coupled map lattice model where the one-body probability 
distribution functions of local (continuous) variables has two disjoint 
supports. By introducing Ising spin variables, the phase ordering process 
following uncorrelated initial conditions was numerically studied and complete 
phase ordering was observed for large coupling values. Both the persistence 
probability $p(t)$ (i.e. the proportion of spins that has not changed sign up to 
time $t$) and the characteristic length of domains $R(t)$ (evaluated as the 
width at midheigth of the two-point correlation function) showed scaling 
behavior and the scaling exponents $z$ and $\theta$ ( defined by the scaling 
laws $R(t)\sim t^{z}$ and $p(t)\sim t^{-\theta}$) were found to vary 
continuously with parameters, at odds with traditional models. The study of the 
phase ordering properties also allowed to determine the limit value $g_c$ beyond 
which multistability disappears and NTCB is observed \cite{lemaitre}. Indeed the 
following relations were found to hold: $\theta\sim(g-g_c)^{w}$ and $z\sim 
(g-g_c)^w$, and were used to determine $g_c$.
The ratio $\theta /z$ was found to be close to $0.40$, the ratio known for the 
time dependent Ginzburg-Landau equation.

Subsequently dynamical scaling was studied in a lattice model of chaotic maps 
where the corresponding Ising spin model conserves the order parameter 
\cite{ang}. This model is equivalent to a conserved Ising model with couplings 
that fluctuate over the same time scale as spin moves, in contact with a thermal 
bath at temperature $T$. The scaling exponents $\theta$ and $z$ were found to 
vary with temperature. In particular the growth exponent $z$ was observed to 
increase with temperature; it follows that thermal noise speeds up the phase 
ordering process in this class of models. At high temperatures $z$ assumes the 
value $1/3$, corresponding to the universality class of a Langevin equation 
known as model $B$ \cite{gunton}, that describes the standard conserved Ising 
model (when bulk diffusion dominate over surface diffusion \cite{huse}).

The role of noise as an ordering agent has been broadly studied in recent years 
in the context of both temporal and spatiotemporal dynamics. In the temporal 
case, that was first considered, external fluctuations were found to produce and 
control transitions (known as {\it noise-induced transitions}) from monostable 
to bistable stationary distributions in a large variety of physical, chemical 
and biological systems \cite{horst}. As far as spatiotemporal systems are 
concerned, the combined effects of the spatial coupling and noise may produce an 
ergodicity breaking of a bistable state, leading to phase transitions between 
spatially homogeneous and heterogeneous phases. Results obtained in this field 
include critical-point shifts in standard models of phase transitions 
\cite{vanden}, pure {\it noise-induced phase transitions} \cite{vanden1}, 
stabilization of propagating fronts \cite{armero}, and noise-driven structures 
in pattern-formation processes \cite{garcia}. In all these cases, the 
qualitative (and somewhat counterintuitive) effect of noise is to enlarge the 
domain of existence of the ordered phase in the parameter space.

It is the purpose of this paper to analyse the role of additive noise in the 
phase separation of multiphase coupled map lattices \cite{losson}. It will be 
shown that external noise can induce complete phase ordering for coupling values 
not leading to phase separation in the absence of the noise term. Furthermore 
this dynamical transition is reentrant: phase separation appears at a critical 
value of the noise intensity but disappears again at one higher value of the 
noise strength. 

The paper is organized as follows. In the next section the coupled map lattice 
model here considered is introduced. In section 3 we present our numerical 
results. Section 4 summarizes our conclusions.

\section{The model}

Let us consider a two-dimensional square lattice of coupled identical maps $f$ 
acting on real variables $x_i$, whose evolution is governed by the difference 
equation:
\begin{equation}
x_i(t+1)=(1-4g)f[x_i(t)]+g\sum_{\langle ij\rangle} f[x_j(t)] +\xi_i (t).
\label{eq1}
\end{equation}
Here the sum is over the nearest neighbors of site $i$, $\xi_i$ is a random 
number uniformly distributed in $[-\sigma /2,\sigma /2]$, $g$ is the coupling 
strength and periodic boundary conditions are assumed. We have chosen the 
following map:
\begin{equation}
f(x)= \left \{
\begin{array}{cl}
 -{\mu\over 3}\exp{[ \alpha(x+{1\over 3})}] & if\;\; x\in[-\infty,-{1\over 3}] 
\\
 \mu x & if\;\; x\in [-{1 \over 3},{1 \over 3}] \\
 {\mu\over 3}\exp{[ \alpha({1\over 3}-x)}] & if\;\; x\in[{1\over 3},+\infty] 
\end{array} 
\right .
\label{map}
\end{equation}
that is defined for every $x$ in the real axis (see Fig. 1). The map here 
considered is a modified version of the map used in \cite{lemaitre}; the 
modification is motivated by the fact that, due to the term noise $\xi_i$, 
variables $x_i(t)$ are not constrained to take value in $[-1,1]$. Choosing $\mu 
=1.9$ and $\alpha =6$, $f$ has two simmetric chaotic attractors, one with $x>0$ 
and the other with $x<0$. In Fig. 2 we show the invariant distribution of the 
attractor with positive $x'$s: it is composed of smooth pieces. The Lyapunov 
exponent of the map was evaluated $0.558$.

To study the phase ordering process, uncorrelated initial conditions were 
generated as follows: one half of the sites were chosen at random and the 
corresponding values of $x$ were assigned according to the invariant 
distribution of the chaotic attractor with $x>0$, while the other sites were 
similarly assigned values with $x<0$. We associated an Ising spin configuration 
$s_i(t)={\rm sgn} [x_i(t)]$ with each configuration of the $x$ variable. Large 
lattices (up to $1000 \times 1000$) with periodic boundary conditions were used; 
the persistence $p(t)$ was measured as the proportion of sites that has not 
changed $s$ the initial value. The average domain size $R(t)$ was measured by 
the relation $C[R(t),t]=1/2$, where $C(r,t)=\langle s_{i+r}(t) s_i(t)\rangle$ is 
the two point correlation function of the spin variables. Both $p(t)$ and $R(t)$ 
were averaged over many (up to thirty) different samples of initial conditions.

\section{Results}

Fixing $\sigma =0$, that is considering the noise-free case, we performed the 
analysis suggested in \cite{lemaitre}. For various values of $g$ we measured the 
characteristic length $R$ and the persistence $p$ as functions of time; both 
these quantities saturate for small couplings and show scaling behaviour for 
large $g$ values. The associated exponents, respectively $z$ and $\theta$, were 
continuous functions of $g$ well described by the fitting ansatz 
\cite{lemaitre}:
\begin{equation}
z\sim (g-g_c)^w\; ,\;\;\;\theta \sim (g-g_c)^w.
\label{eqfit}
\end{equation}
The estimated values of $g_c$ and $w$ were $g_c=0.1652$ and $w=0.2260$ while 
fitting the exponent $z$, and $g_c=0.1654$, $w=0.2105$ for the exponent 
$\theta$. The ratio $\theta /z$ was approximately independent of $g$ and equal 
to $0.3767$. Furthermore, we observed that the same fitting ansatz can be used 
to fit our data for nonvanishing and small noise strength $\sigma$. For example, 
in Fig. 3(a) and 3(b) we show respectively the fit of $z$ and $\theta$ versus 
$g$, while 
keeping $\sigma $ fixed and equal to $0.1$. As one can see, data are well fitted 
by the scaling forms (\ref{eqfit}), and the estimated values are $g_c=0.1628$, 
$w=0.2197$ for the $z$ exponent, and $g_c=0.1632$, $w=0.2024$ for the $\theta$ 
exponent \cite{oss}. The ratio $\theta /z$ was estimated at $0.3838$. We remark 
that our estimate of the critical coupling $g_c$, when non-vanishing and small 
noise is present, is smaller than the noise-free critical value. This fact 
clearly shows that a proper amount of noise favours the phase separation process 
of the system. 

Let us now consider the region $g < g_c(\sigma =0) = 0.165$. Here in the 
noise-free case the system evolves towards blocked configurations and no phase 
separation takes place. We checked, however, that this asymptotic regime was 
attained after very long evolution times: the system spended a lot of time in 
metastable states, so that the evolution curve for $R$ and $p$ displayed typical 
stairs structure. This structure (the times marking the steps of the curve) was 
very robust, in the sense that:
\renewcommand{\labelitemi}{$-$}
\begin{itemize}
\item it resisted to a change of the initial conditions (choosen following the 
particular prescription of section II),
\item it did not depend on lattice dimension,
\item a little noise (low $\sigma$) did not destroy it.
\end{itemize}
Nevertheless, when growing the amount of noise, the life time of these 
metastable states became shorter and shorter, till they definitely disappeared 
for $\sigma$ greater than a critical value $\sigma_c(g)$. For 
$\sigma>\sigma_c(g)$ we got again power laws for $R(t)$ and $p(t)$, showing that 
the system separates for large times. This behaviour is shown in Fig. 4.

We estimated the critical value $\sigma_c$ by fitting our data with the ansatz 
$z\sim (\sigma -\sigma_c )^w$. In Fig. 5 we show our data corresponding to 
$g=0.16$: we evaluated $\sigma_c=0.1094$ and $w=0.3152$. As in the case of the 
choice \ref{eqfit}, we have no theoretical argument to support the choice of the 
fitting ansantz, but the fact that it works on a large interval of $g$ letting 
us to give a precise measurement of $\sigma_c$. We were able to measure in such 
a way $\sigma_c$ for $g$ greater than $0.025$; at smaller values of $g$ the 
dynamics became very slow and we were not able to numerically extract the 
exponent $z$.

As $\sigma$ was increased, we found a transition at another critical value of 
the noise strength showing that the system does not separate beyond this 
critical $\sigma$. As an example in Fig. 6 we show the exponent $z$ versus 
$\sigma$ for $g$ fixed and equal to $0.17$. The transition seems to be 
discontinuous. 

We repeated this analysis for several values of $g$. Interpolating the above 
described data for the critical noise strength, we built the phase diagram for 
the model shown in Fig. 7. The system separates in the shaded area, that is it 
tends asymptotically to complete phase ordering. Points in the white area 
correspond to an asymptotic regime of the system where clusters of the two 
phases are dynamical but their mean size remains constant; only for $\sigma =0$ 
one has blocked configurations with clusters fixed in time. Our data concern $g$ 
greater than 0.025, however we extrapolated the two critical curves towards 
$g=0$.  We observe, interestingly, that the extrapolation of the two curves seem 
to meet at $g=0$; further investigation is needed to clarify the behavior of the 
noisy system close to $g=0$.

\section{Conclusions}

The phase ordering properties of multiphase chaotic map lattices have recently 
attracted interest since they differ from those of traditional models. In this 
paper we have shown that additive noise acts as an ordering agent in this class 
of systems, i.e. for a suitable amount of noise the system may order even for 
values of the coupling strength for which no separation is observed in the 
absence of the noise-term. A simple explanation for this behavior is as follows. 
Small values of the spatial coupling lead, in the noise-free case, to spatially 
blocked configurations where interfaces between clusters of each phase are 
strictly pinned. A proper amount of noise makes the system cross these barriers 
thus leading to complete ordering. We have numerically constructed a phase 
diagram describing this behavior. As we said, a similar effect  was observed in 
chaotic map lattices evolving with conserved dynamics, where we found that the 
growth exponent increases with temperature \cite{ang}; in the present case the 
additive noise plays the role of the thermal noise.

\newpage
\noindent\Large\textbf{Figure Captions}
\normalsize
\vspace{1.0cm}
\begin{description}
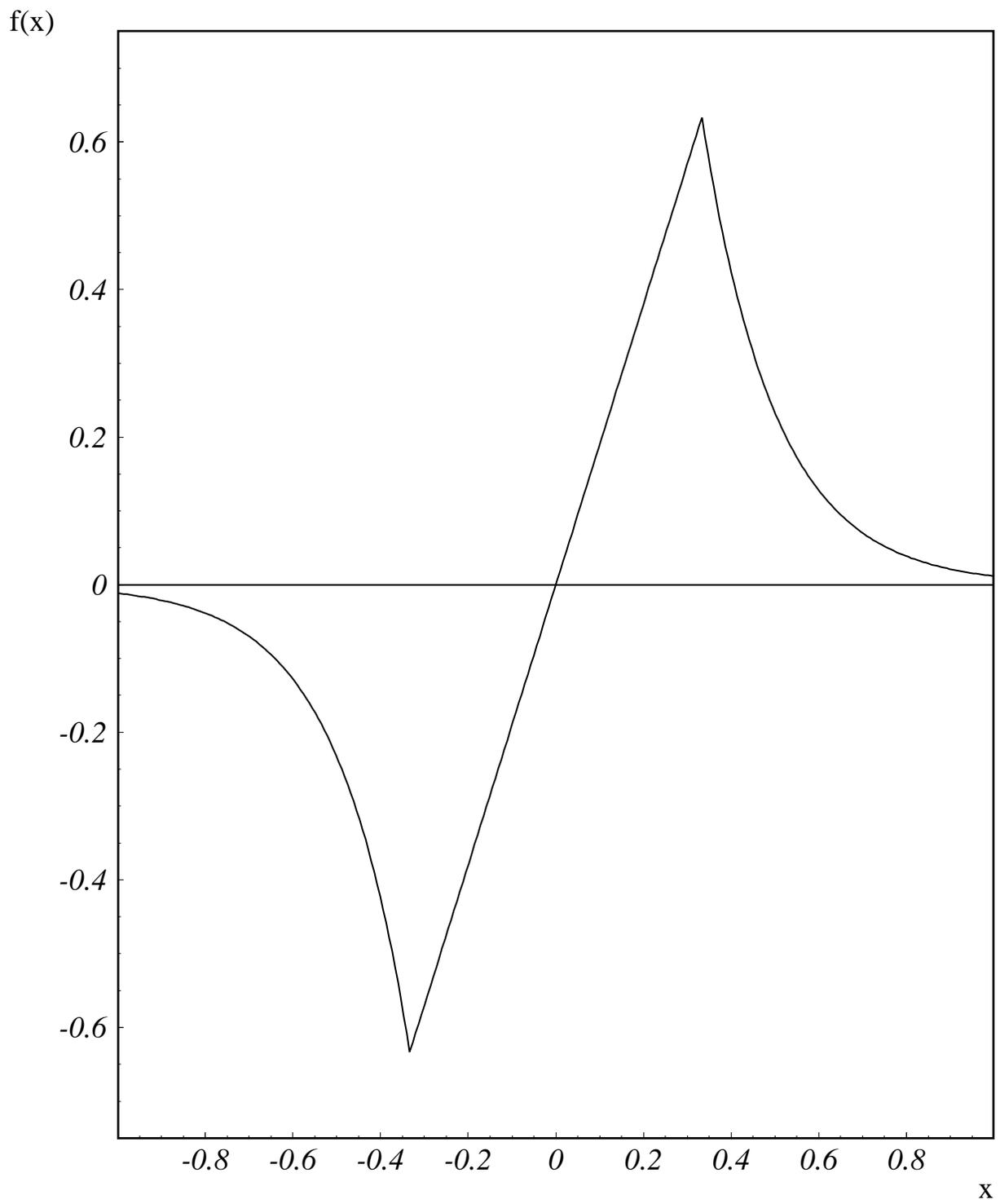
\item{Figure 1}: The map $f(x)$ defined in (\ref{map}).
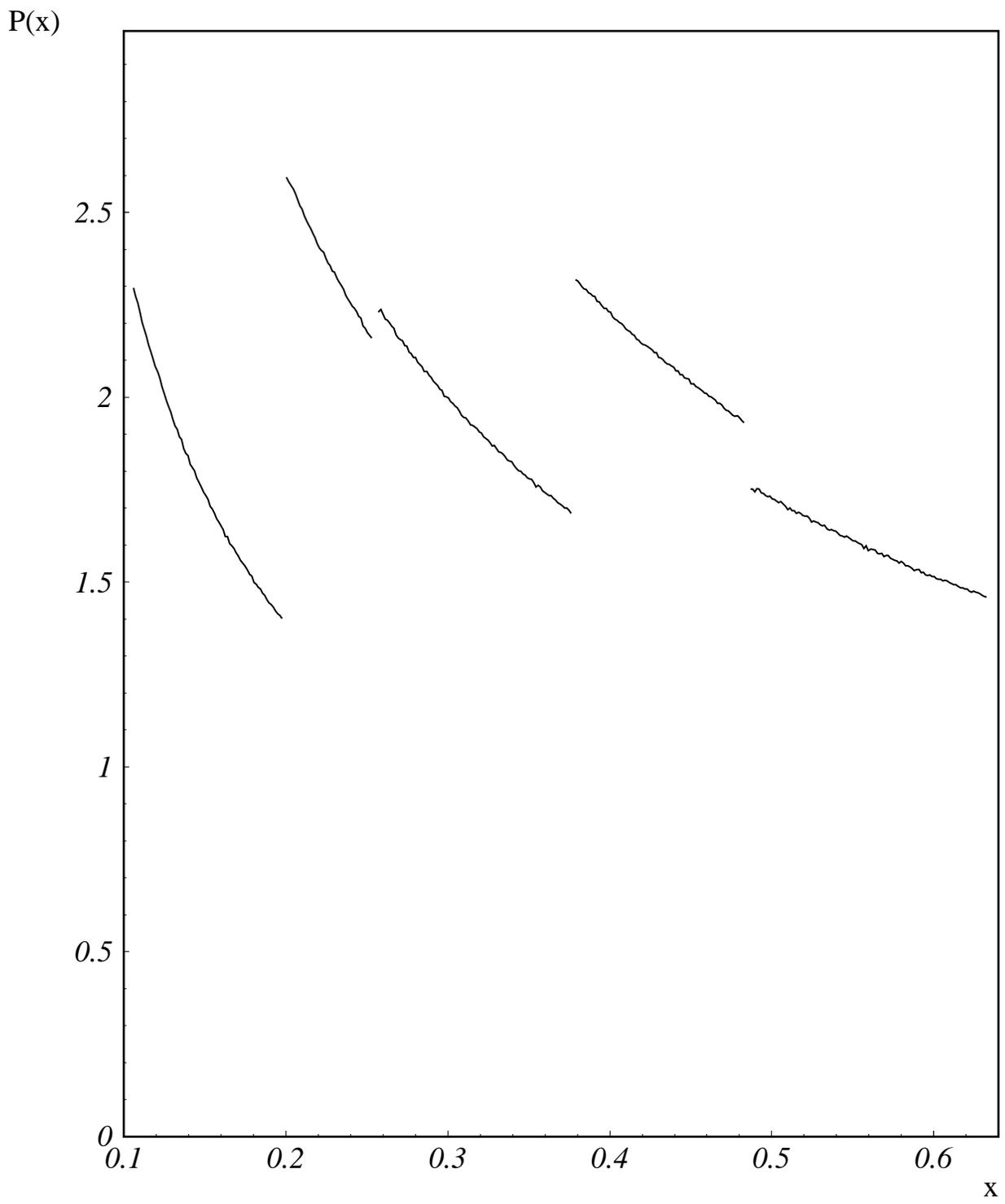
\item{Figure 2}: Invariant probability distribution for the positive attractor 
of $f(x)$.
\item{Figure 3}: The estimated scaling exponents at fixed noise $\sigma=0.1$ : 
a) the dependence of the growth exponent $z$ from $g$ in linear and log-log 
plot, b) the dependence of the persistence exponent $\theta$ from $g$ in linear 
and log-log scale. Solid lines are best fits leading to the determination of 
$g_c$ and $w$ through the use of (\ref{eqfit}).
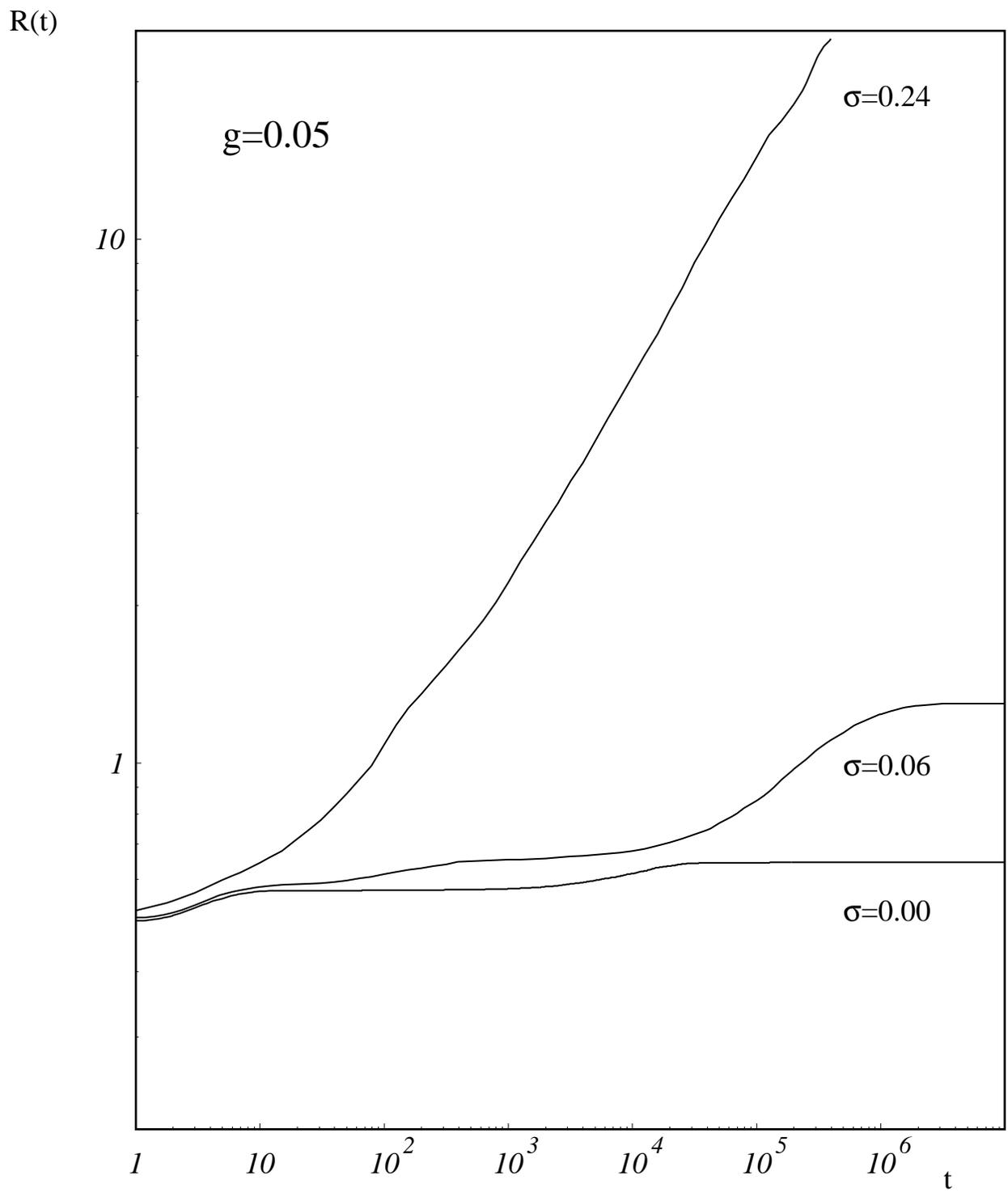
\item{Figure 4}: The effect of additive noise on the time evolution of the 
domain size $R(t)$ at $g=0.05$. The three curve are relative to $\sigma=0, \sigma=0.06, 
\sigma=0.24$.
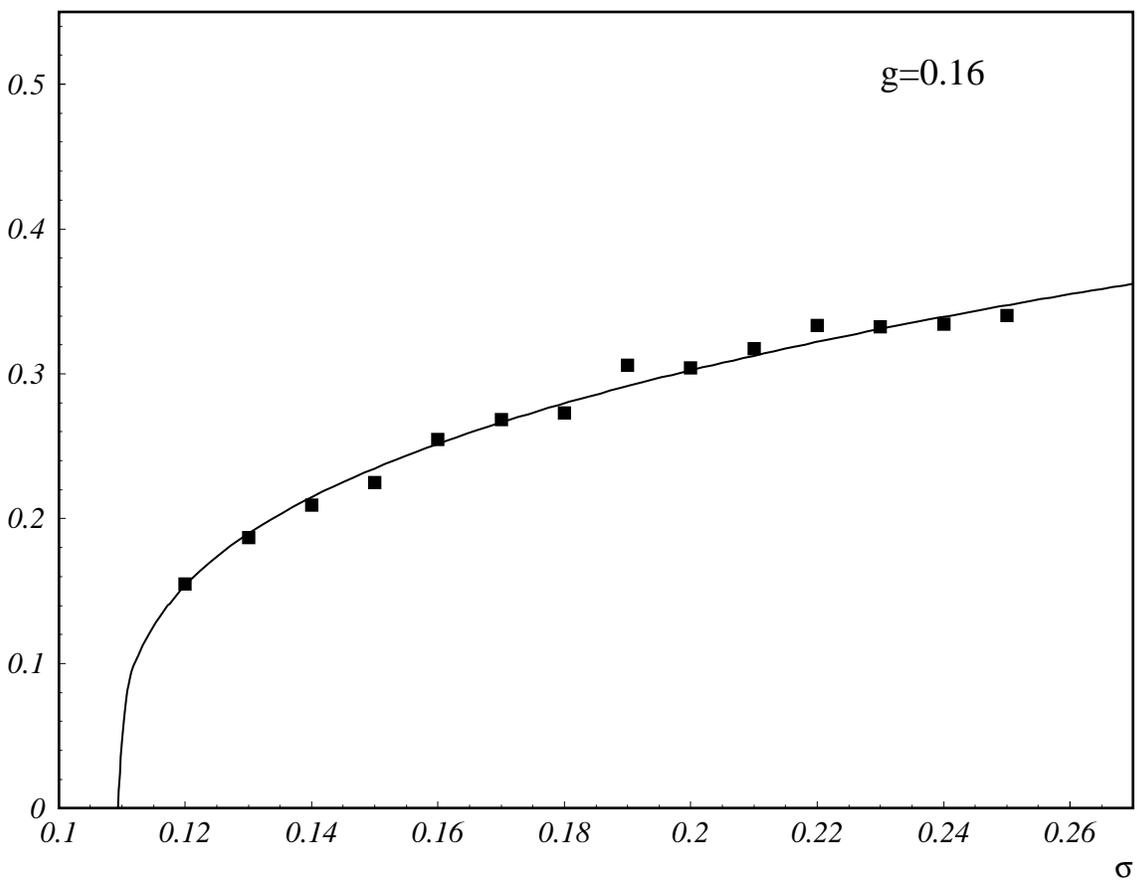
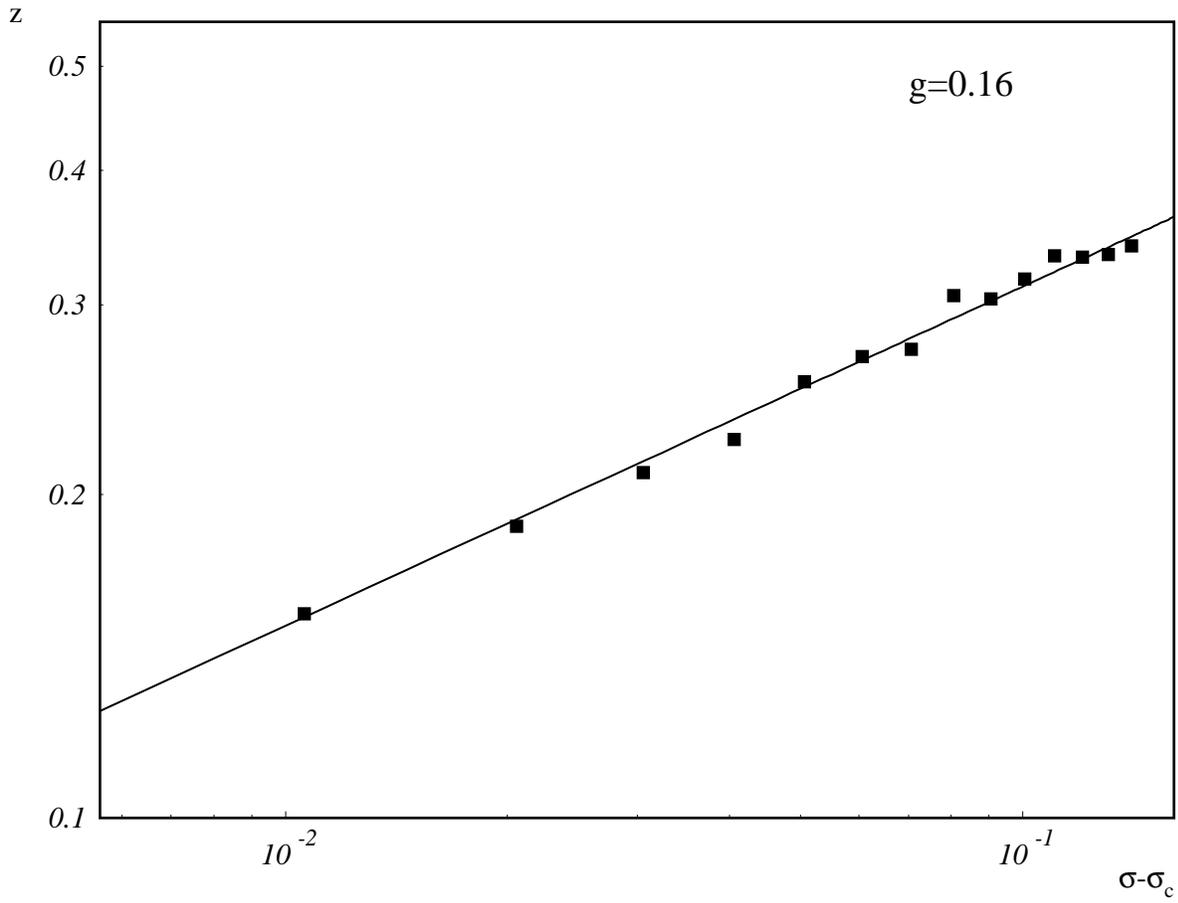
\item{Figure 5}: The estimated growth exponent $z$ versus $\sigma$ at fixed 
coupling $g=0.16$ in linear and log-log scale. Also shown is the best fit with 
the function $z\sim (\sigma -\sigma_c )^w$.
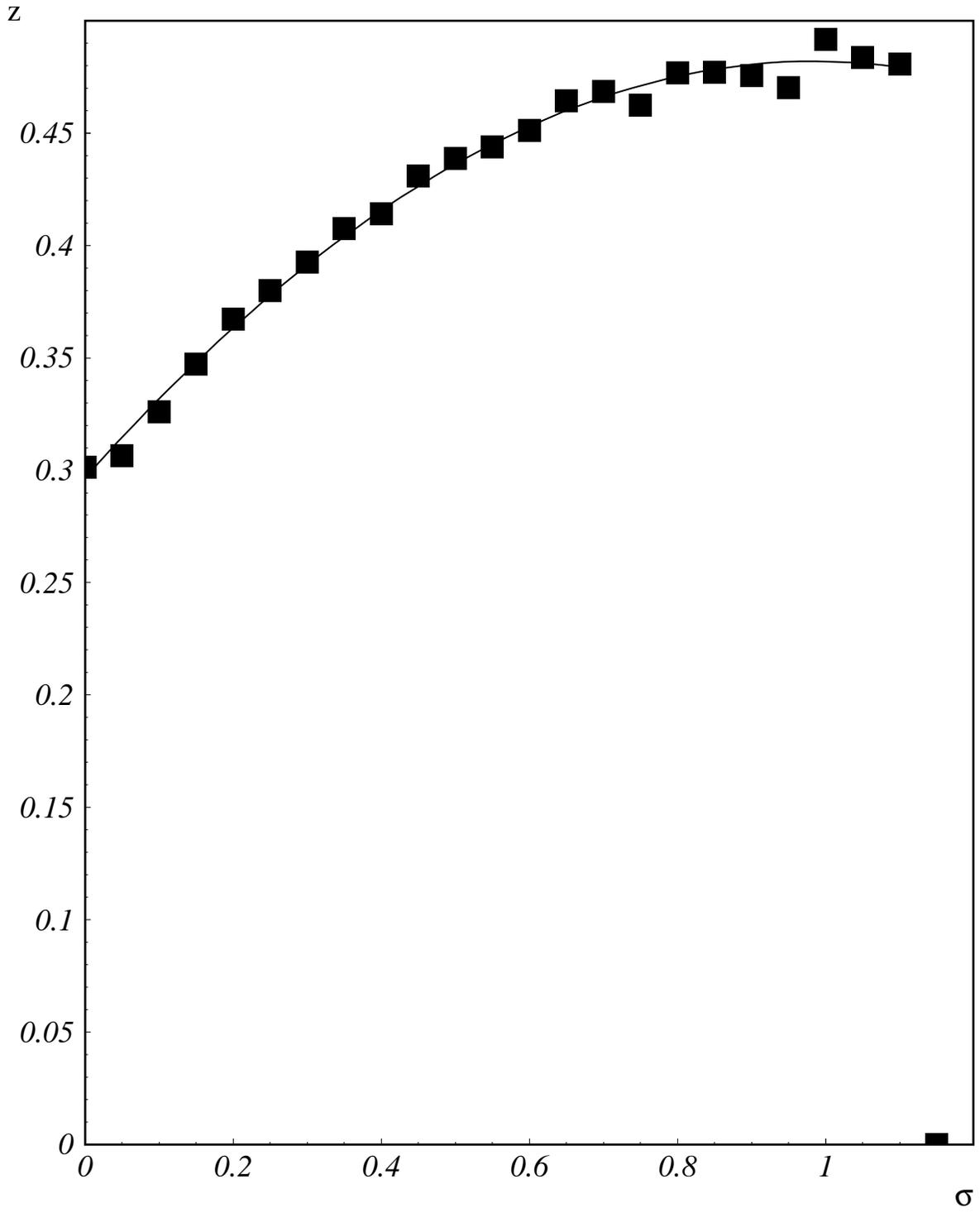
\item{Figure 6}: The estimated growth exponent $z$ versus $\sigma$ at fixed 
coupling $g=0.17$. $z$ goes abruptly to zero at $\sigma=1.2$ showing that the 
system does not separate beyond this threshold.
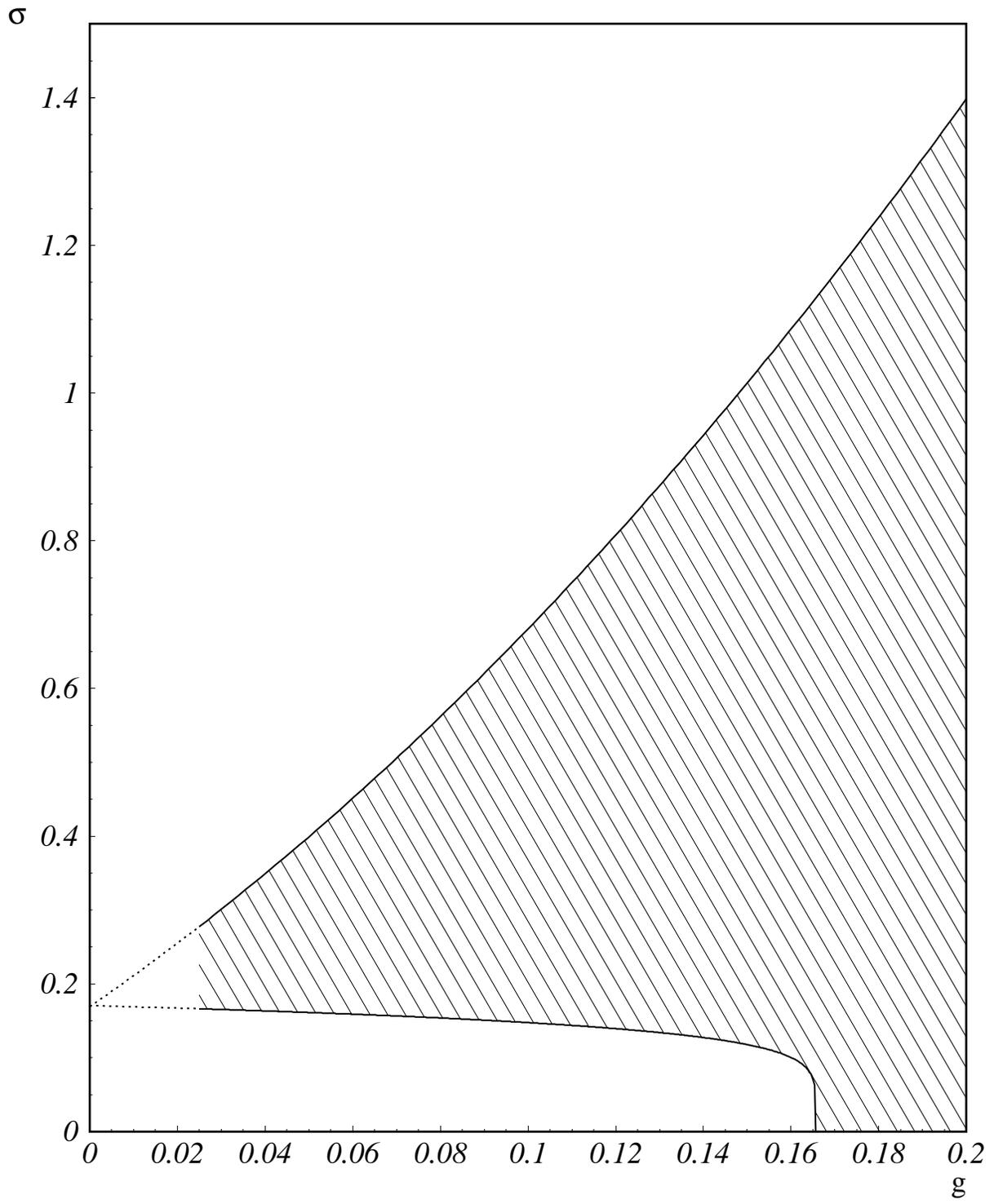
\item{Figure 7}: The phase diagram in the plane $\sigma-g$. The shaded area 
represents the parameter region in which the system separates asymptotically.

\end{description}
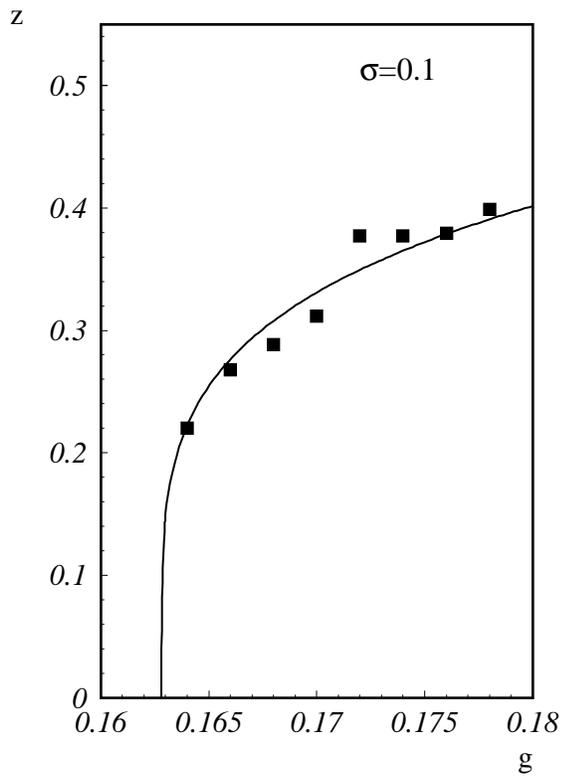
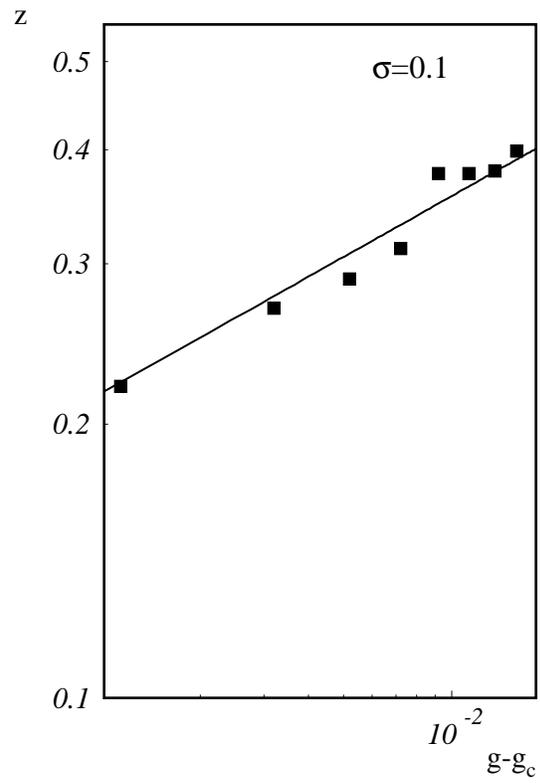
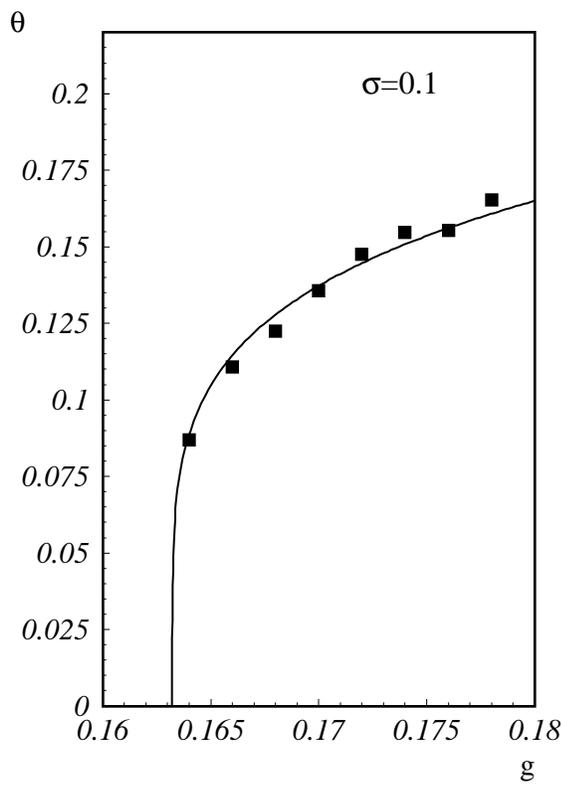
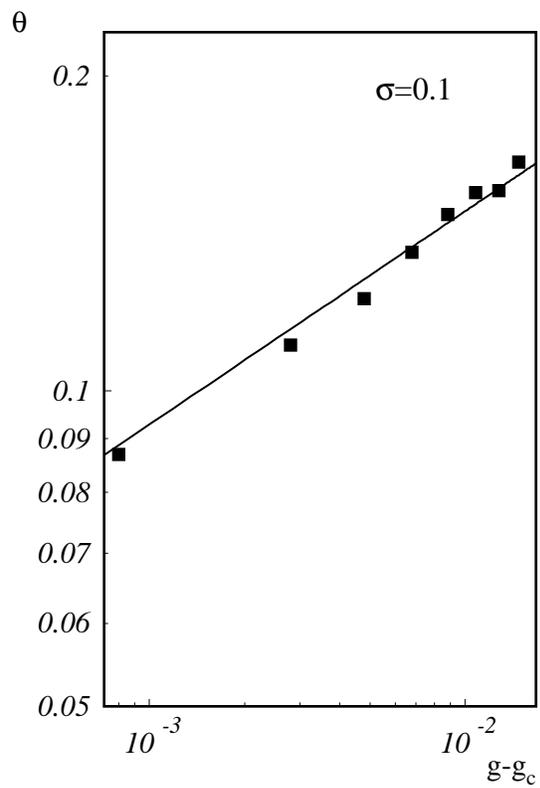
\end{document}